# Non-volatile Electric Control of Magnetic and Topological Properties of MnBi$_2$Te$_4$ Thin Films

Wei Luo[1], Mao-Hua Du[2], Fernando A. Reboredo[2], Mina Yoon[2]*

[1]Center for Nanophase Materials Sciences, Oak Ridge National Laboratory, Oak Ridge, TN 37831, USA

[2]Materials Science and Technology Division, Oak Ridge National Laboratory, Oak Ridge, TN, 37831, USA

* Author to whom any correspondence should be addressed.
E-mail: myoon@ornl.gov

## Abstract

In this letter, we propose a mechanism to control the magnetic properties of topological quantum material (TQM) by using magnetoelectric coupling: this mechanism uses a heterostructure of TQM with two-dimensional (2D) ferroelectric material, which can dynamically control the magnetic order by changing the polarization of the ferroelectric material and induce possible topological phase transitions. This concept is demonstrated using the example of the bilayer $MnBi_2Te_4$ on ferroelectric $In_2Se_3$ or $In_2Te_3$, where the polarization direction of the 2D ferroelectrics determines the interfacial band alignment and consequently the direction of the charge transfer. This charge transfer, in turn, enhances the stability of the ferromagnetic state of $MnBi_2Te_4$ and leads to a possible topological phase transition between the quantum anomalous Hall (QAH) effect and the zero plateau QAH. Our work provides a route to dynamically alter the magnetic ordering of TQMs and could lead to the discovery of new multifunctional topological heterostructures.

1. **Introduction**

The interplay between topology and magnetism can lead to many exotic states, such as magnetic Weyl semimetals [1-10], quantum anomalous Hall (QAH) effect [11-17], axion insulators [18,19], higher order topological insulators [20-22], and topological superconductors[23]. Recently, $MnBi_2Te_4$ (MBT) and MBT/$(Bi_2Te_3)_n$ have attracted attention because of their intrinsic magnetism and topological band structures [16,24-30]. However, these materials show experimentally an antiferromagnetic (AFM) ground state between different layers, which requires a large magnetic field to align the spin direction and realize the QAH insulator [16,31], limiting their application.

To realize the QAH effect in a low magnetic field, some efforts have been made to tune the AFM to ferromagnetic coupling between layers. Chemical doping is one of the most studied methods to control magnetic coupling. For example, recent studies have shown that ferromagnetism (FM) can arise in Mn(Bi1-xSbx)$_4$Te$_7$ with x = 0.3 [32], and the p-type dopants can induce the AFM-to-FM transition and realize the high-temperature QAH effect [33]. Another effective method is to use site mixing (antisite defects Mn$_{(Bi,Sb)}$) to enhance the FM or change the magnetic states [34,35]. All of these methods require chemical doping, which changes defect chemistry and in some cases can also significantly increase the defect density and disorder (e.g., in the case of Sb doping) [36].

In this letter, we propose a new strategy to enhance FM in MBT on a substrate. Our proposed system consists of a heterostructure of MBT on a 2D ferroelectric material, where the interface-charge transfer between the subsystems could be dynamically controlled by changing the polarization of the ferroelectric substrate with an applied electric field. We reveal magnetoelectric coupling in which the charge transfer causes a change in the magnetic order of the MBT as well as its quantum topological state. Our calculations show that p-type electronic doping is more effective in promoting FM in MBT; moreover, the p- or n-type electronic doping and the possible topological phase transition between the QAH and the zero-plateau QAH (ZPQAH) phases in bilayer MBT can be controlled by switching the polarization of a ferroelectric substrate In$_2$Se$_3$ or In$_2$Te$_3$.

## 2. Computational detail

All calculations are based on density functional theory (DFT) [37,38] implemented in the VASP code [39]. The interaction between ions and electrons is described by the projector augmented wave method [40]. For bulk MBT, the total energy was calculated using the Perdew-Burke-Eznerhof exchange correlation functional [41] and a kinetic energy cutoff of 270 eV. The convergence of the cutoff energy is tested against the energy difference between the AFM and FM ($\delta E_{AFM-FM}$) in bulk MBT with zero carrier density. The result shows that increasing the cutoff energy from 270 to 400 eV leads to an increase of $\delta E_{AFM-FM}$ by only 0.03 meV/Mn. A 3 × 3 × 2 supercell, which contains six single layers (SLs), was used for defect calculations in bulk MBT (see Figure. 1a). Lattice parameters are optimized, and atomic positions were relaxed until the forces were less than 0.01 eV/Å. The optimized lattice constants of bulk MBT, a = 4.355 Å and c = 40.596 Å, are in good agreement with experimental values of a = 4.3338 Å and c = 40.931 Å [42]. We then construct a heterostructure consisting of a bilayer MBT on an SL $In_2Se(Te)_3$, where the kinetic energy limit is 500 eV and the Brillouin zone (BZ) is sampled with a 15 × 15 × 1 k-mesh. We employ DFT-D3 vdW functional [43] (which is proved to be appropriate for describing the van der Waals interaction for an MBT system [44]) and make a dipole correction. A $U$ parameter of 3 eV is applied to Mn 3d orbitals [45] for both bulk MBT [24] and the heterostructure calculations. In addition, we test our results for $U$ value ranges from 1eV to 4eV (since the U=3eV is widely used for MBT systems). In our calculations, the strain for bilayer MBT is zero since we claim that the enhancement of FM state is purely caused by charge transfer instead of by the strain effects from substrates. In fact, the interaction between bilayer MBT and $In_2Se(Te)_3$ is van der Waal interaction. Thus, the lattice constant of MBT cannot be affected much by the substrate $In_2Se(Te)_3$. The strains for the substrate $In_2Se_3$ and $In_2Te_3$ are 5.8% and -1.1%, respectively. This a bit large strain 5.8% for $In_2Se_3$ indicates that we need a supercell 18×18 bilayer MBT/19×9 $In_2Se_3$ to eliminate the strain effects for $In_2Se_3$. This is not feasible from DFT calculations. For the simplicity of calculations and discussions, the size of the supercell 1×1 bilayer MBT/1×1 $In_2Se(Te)_3$ is used. This is reasonable since the ferroelectric polarization of $In_2Se(Te)_3$ changes a little under a biaxial strain[46]. On the other hand, from the perspective of experiment, for example, the molecular beam epitaxy, the strain is nature come from the upper component of the heterostructure, a reduced strain in the system is also to demonstrate the feasibility of the model heterostructure. The optimized lattice constants of free-standing bilayer MBT, $In_2Se_3$ and $In_2Te_3$ are about 4.346 Å, 4.106 Å and 4.397 Å. We use Wanniertools[47,48] to calculate Chern number and surface states based on the Wilson loop method[49] and Green's functions[50].

**Free Carriers Enhance FM in Bulk MBT**

Carrier-mediated FM has been extensively studied in diluted magnetic semiconductors, in which the moments of magnetic dopants can be ferromagnetically aligned by spin-dependent coupling with free holes or electrons [51-54]. We show here that a similar mechanism can promote FM in MBT. Both the valence and conduction bands of MBT are made up of *p*-orbitals of Te and Bi, which have coupling with Mn-3*d* orbitals. The *p-d* exchange coupling between the conduction/valence band states and Mn-3*d* states can give rise to the spin-splitting of the band states. The introduction of free holes/electrons into the spin-split valence/conduction band enhances FM. Note that the mechanism here is different from the hole-doping induced ferromagnetism in group III 2D monochalcogenides[55-58]. In group III 2D monochalcogenides, the density of states near the valence bands exhibits a sharp van Hove singularity. Thus, doping with hole could induce a ferromagnetic state. On the other hand, the strong spin-orbit coupling (SOC) in MBT mixes up spin-up and -down electrons, thereby suppressing the carrier-mediated FM. Since the SOC effect is stronger in Bi-6*p* orbitals than in Te-5*p* orbitals, the spin-splitting of the conduction band states, which are primarily made up of Bi-6*p* orbitals (see Figure. S1 of supporting information [SI]), should be much weaker than that of the valence band states, which are dominated by Te-5*p* orbitals (see Figure. S1 of SI). Therefore, the free hole is expected to be more effective in promoting FM in MBT than the free electron. This is indeed confirmed by our calculations as shown subsequently.

We investigated the effects of free carriers on magnetism in bulk MBT by calculating the energy difference between the AFM and FM orderings as a function of the free carrier density as shown in Figure. 1b. Free carriers were introduced by directly adding electrons or holes in pristine MBT (red squares in Figure. 1a) or by creating shallow donor or acceptor defects in a MBT supercell (blue circles in Figure. 1a). $Bi^+_{Mn}$ and $Mn^-_{Bi}$ antisite defects (Figure. 1c) have previously been identified as the primary shallow donor and acceptor defects, respectively, in MBT [36]; therefore, they were introduced to generate free electrons and holes. Pairs of $Bi^+_{Mn}$ and $Mn^-_{Bi}$ antisite defects were incorporated in MBT SLs in the case of zero free carrier density in Figure. 1c, and the relative concentration between $Bi^+_{Mn}$ and $Mn^-_{Bi}$ were modified to produce a net free electron or hole density.

As can be seen in Figure. 1a, the two approaches of creating free carriers lead to roughly the same trend in the relative stability between AFM and FM (i.e., both the hole and electron can enhance FM). The AFM ordering is more stable than the FM ordering when there is no free carrier. Increasing the hole density promotes FM, which becomes more stable than AFM when the hole density exceeds ~3.5 × $10^{20}$ cm$^{-3}$. This is akin to the hole-mediated FM in diluted magnetic semiconductors [51]. On the other hand, increasing the electron density has a much smaller effect on enhancing FM compared with increasing the hole density. This is because of the SOC effect described previously. Indeed, our calculations show that free electrons also enhance FM significantly if the SOC is not included, as shown in Figure. 1a.

Note that the magnetic moment of a $Mn^-_{Bi}$ defect is aligned parallelly or antiparallelly (see Figure 1c) to the moment of its adjacent native Mn ($Mn_{Mn}$) in FM and AFM MBT depending on the carrier density, as shown in Figure. 1b. Without free carriers, the magnetic moments of $Mn^-_{Bi}$ and $Mn_{Mn}$ are aligned antiparallelly. The introduction of free holes enhances the stability of parallel alignment and eventually flips the spin on $Mn_{Bi}^-$ as the free hole density increases. On the other hand, the free electrons do not promote the parallel alignment of the magnetic moments of $Mn_{Bi}^-$ and $Mn_{Mn}$. This trend can also be understood by the effects of free carriers and the SOC on magnetism as described previously. Bulk MBT is typically n-doped by excess $Bi_{Mn}^+$ [59]. The magnetic moment of a $Mn_{Bi}^-$ defect should be aligned antiparallelly to $Mn_{Mn}$ in n-type MBT, as shown in Figure. 1c. Since $Mn_{Bi}^-$ is abundant in MBT [36] the antiparallel alignment between $Mn_{Bi}^-$ and $Mn_{Mn}$ can explain the relatively small Mn saturation moment of ~3.56 $\mu_B$/Mn (2 K) observed in MBT[42].

## 3. Design Principles for Nonvolatile Carrier Injection

Our results show a promising way to tune the magnetic properties of MBT by controlling the free carrier density in the system, which can be realized by chemical doping. However, introducing free carriers via chemical doping is usually not easy to achieve in experiments. Our alternative way to realize this mechanism in a nonvolatile manner is to use a heterostructure of MBT on a substrate, where electronic doping of MBT can be achieved through the charge transfer interaction with the substrate. To realize the electron (hole) doping for the host material, the electron should transfer from the substrate (host material) to the host material (substrate). This can be realized if the conduction band minimum (CBM) (valence band maximum [VBM]) of the host material is lower (higher) than the VBM (CBM) of the substrate. A desirable band alignment is therefore a "broken-gap" type-III band alignment, where the transfer of electrons or holes is possible by changing the relative location between the CBMs and VBMs of the host material and the substrate, which are labeled cases in Figure. 2a Case I (II) means that the VBM (CBM) of substrate is higher (lower) than the CBM (VBM) of host material, which ensures doping of the host material by electron (hole) carriers.

## 4. Van der Waals Heterostructure between MBT and In$_2$Se(Te)$_3$

### 4.1. SL MBT on In$_2$Se(Te)$_3$

We propose a new mechanism to control the electronic properties of MBT by an external electric field and ultimately control its magnetic ordering. We implement the idea in a heterostructure of MBT on a 2D ferroelectric material that maintains intrinsic polarization [46,60-66], where an electric field can switch the polarization if the barrier is not very large. Changes in polarization alter the energy level of the 2D ferroelectrics relative to that of MBT, leading to transfer of electron or holes carriers into MBT (see the illustration of the mechanism in Figure. 2b). Note that the magnetoelectric coupling here is different from the conventional magnetoelectric coupling, in which the ferroelectricity is caused by the spin-order [67-69]. The idea here is more likely the ferroelectric field-effects [70]. Traditional magnetoelectric multiferroic research has focused on bulk materials in which the lattice and the magnetic degrees of freedom are coupled because of spin-orbit interaction [71]. In this case, in contrast, the coupling between the structure and magnetism occurs at the interface of an heterostructure. Although our calculations include the SOC for accuracy, this is not the primary reason for the magnetoelectric coupling. Instead, the change of the charge transfer direction at the interface is associated with a drastic change in the band alignment due to a structural phase transition.

The heterostructure between 2D ferroelectrics and host materials has been proposed for engineering topological properties [72,73], magnetism [44,74,75] and crystal Hall effect [76]. Here, we consider the heterostructure between MBT and 2D ferroelectrics In$_2$Se(Te)$_3$. To identify the lowest energy stacking configuration between these two van der Waals materials, for simplicity, we first investigated the heterostructure between SL MBT and In$_2$Se(Te)$_3$. We consider three different stacking configurations, denoted here as, hollow1, hollow2, and top configurations (see Figure. S2 of SI). The total energies of the three configurations were calculated including the van der Waals interactions, the SOC, and the dipole correction [77]. Our calculations showed that the hollow1 configuration is the ground state, which is consistent with previous calculations [44]. In addition, we also checked their total energies by testing different Hubbard U values (see Table S1 of SI) and found that the ground state (hollow1 configuration) is robust for different U values. In the following calculations, we focus only on the hollow1 stacking configuration and fixed U value equal to 3 eV for *d* orbital of Mn [24]. Since the FM state of SL MBT is very robust, it is difficult to change its magnetic state. In addition, the SL MBT is topologically trivial with a sizeable band gap [25,26,29], and the In$_2$Se(Te)$_3$ substrate cannot induce a topological phase transition (details analysis can be found in part I of the SI). Therefore, we further investigated the effects of the In$_2$Se(Te)$_3$ substrate on the magnetic and topological properties of a MBT bilayer in the rest of our work.

**4.2. Possible Topological Phase Transition Controlled by the Electric Field for Bilayer MBT on In$_2$Se(Te)$_3$**

Here, we consider a bilayer MBT on a 2D In$_2$Se$_3$ or In$_2$Te$_3$ substrate. 2D In$_2$Se$_3$ or In$_2$Te$_3$ are both ferroelectrics, and the switch of their polarization is associated with a structural phase transition as shown in Figures. 3a and 3d. Due the asymmetry of In$_2$Se$_3$, there is a dipole moment along the +z direction (for Figure 3(a)). If we apply an electric field along the -z direction, the dipole moment will feel a force and start to reverse its direction. The reverse of dipole moment is associated with the moving of the middle Se atoms, leading to a structural phase transition[46]. We show that changing the polarization of the ferroelectric substrate leads to hole or electron doping of the bilayer MBT, which in turn can induce changes in the interlayer magnetic coupling and possible topological phase transitions. First, we investigate the energy difference between AFM and FM (DE=$E_{AFM}$-$E_{FM}$) order between the bilayer MBTs on $In_2Se(Te)_3$ (see Figure. S5 of SI) and find the polarization of the substrate can drastically change the energy difference. For example, switching the polarization direction of $In_2Se_3$ from the "up" to "down" state (Figure. S5 of SI) changes DE from -1.1 to -0.2 meV/Mn, enhancing the FM ordering significantly. In contrast, if the substrate is $In_2Te_3$, the AFM state of MBT is much more stable (DE = -1.42 meV/Mn with the down polarization) and the change of polarization improves the relative stability of FM by 0.46meV/Mn (i.e., DE for the up polarization is -0.96 meV/Mn). In addition, the energy differences between AFM and FM for different polarization states are checked for different U values, and these results are robust for different U values (see Table S2 of SI).

To understand why the substrate can promote the FM ordering in MBT, we investigated the electronic structure of the bilayer MBT/$In_2Se(Te)_3$ heterostructure, especially the band alignment after contacting. To be specific, for $In_2Se_3$ substrate, if the polarization is up, the ground state of the bilayer MBT is AFM (Figure. 3a). Its electronic band structures with and without SOC effects are shown in Figure. 3b and Figure. 3c (For their density of states, please see Figure. S6 of SI). Without SOC effects, there is no band inversion at Γ point between p orbitals of Bi (blue circles) and p orbital of Te (green circles), because the conduction bands are purely contributed by p orbitals of Bi (neglect the bands from $In_2Se_3$) and the valence bands are purely contributed by p orbitals of Te. After SOC included, the band inversion emerges at Γ point [26] since the conduction bands have some contribution from $p$ orbitals of Te, and the valence bands have some contribution from $p$ orbitals of Bi. The non-trivial band gap is about 76 meV (see the embedded figure in Figure.

3c). Although there is a band inversion, the Chern number of this system is 0 and maintains gapped surface states (see Figure. S7 of SI) since the system is an AFM state. Previous study named this kind of state ZPQAH state since it maintains the flat regions in the [26] hysteresis-like dependence of the Hall conductivity on the external field $\sigma_{xy}(H)$. Within a certain range of the external magnetic field H (smaller than the coercivity), the Hall conductivity is $\sigma_{xy}(H) = 0$. When the H is larger than the coercivity, the Hall conductivity becomes $\sigma_{xy}(H) = +\frac{e^2}{\hbar}$ or $\sigma_{xy}(H) = -\frac{e^2}{\hbar}$, depending on the magnetization direction. This unique feature makes the ZPQAH a good platform for topological spintronics.

As we reverse the polarization of the In$_2$Se$_3$ substrate with an electric field [46], the FM state for the bilayer MBT will be enhanced and the AFM has just 0.2 meV/Mn lower energy than the FM state (the small energy difference means that it is easy to align the spin to FM [Figure. 3d] with a magnetic field). Here, we investigate the electronic properties for the FM state. Without SOC effects, there is no band inversion between p orbital of Bi and p orbital of Te. However, there exists a band inversion between s orbital of In and p orbital of Te (Figure. 4e). Note that such band inversion between the s orbital of In and the p orbital of Te is caused by the band alignment instead of SOC; thus, it is not related with the topological nontrivial properties. After SOC is included, the band inversion happens between the p orbital of Bi and the p orbital of Te and the band gap is about 38 meV (see embedded figure in Figure. 3f), combining with FM state (note that the magnetic moments are along the z direction here), tuning the system to the QAH state (see Figure. S7 of SI). Since the Chern number is strongly related to the direction of the magnetic moments, we check the magnetic anisotropic energy (MAE) as a function of the U values (1-4eV) (see Table S3 of SI) and confirm that the easy axis along the z direction (corresponding to the QAH state) is preferred regardless of the U values. The band inversion between the s orbital of In and the p orbital of Te is caused by the band alignment instead of SOC; thus, it is not related with the topological nontrivial properties.

For the In$_2$Te$_3$ substrate, if the polarization is down, the system maintains an AFM ground state (Figure. 4a). Comparing Figure. 4b with Figure. 4c (For their density of states, please see Figure. S6 of SI), one can find that the SOC effects induce the band inversion between the p orbital of Te (from bilayer MBT) and the p orbital of Bi. The system is in ZPQAH state, and the band gap is 83 meV. If the polarization of In$_2$Te$_3$ is reversed (considering the FM state, Figure. 4d), without SOC, the p orbitals of Te (from In$_2$Te$_3$, yellow circles) are lifted compared with the polarization-down case. With SOC, a band inversion between the p orbital of Te (from bilayer MBT) and the p orbital of Bi appears at Γ point, combining the FM state, tuning the system to the QAH state with a 27 meV band gap. Hence, for both In$_2$Se$_3$ and In$_2$Te$_3$ substates, because of the magnetoelectric coupling with MBT (through hole and electron doping, for each), the FM state can be enhanced, and possible topological phase transitions can be induced by different polarization states, which can be controlled by an electric field.

## 5. Band Alignment for Bilayer MBT on In$_2$Se(Te)$_3$

The preceding band structure calculations indicate that the band alignment between In$_2$Se$_3$ and In$_2$Te$_3$ with MBT is very different from changes in electron doping; thus, magnetic stabilities of the MBT depend on the polarization direction. As an illustration, Figure. 5 shows the band alignment after contacting (We also calculate the band alignment before contacting and the work functions for free standing In$_2$Se(Te)$_3$ and bilayer MBT, see Figure S8 of SI) for four different cases: (1) polarization-down with FM, (2) polarization-up with AFM for bilayer MBT on In$_2$Se$_3$, (3) polarization-up with FM, and (4) polarization-down with AFM for bilayer MBT on In$_2$Te$_3$. For each case, we chose the CBM of bilayer MBT as the reference point (zero-energy). One can see that, for the $In_2Se_3$ substrate, with polarization-down state (FM), the electron can transfer from the bilayer MBT to $In_2Se_3$, leading to the hole doping for bilayer MBT. With the polarization-up state (AFM), the electron transfer is not possible because of the type-I band alignment. For the $In_2Te_3$ substate, with the polarization-up state (FM), the electron can transfer from $In_2Te_3$ to the bilayer MBT, leading to the electron doping at bilayer MBT. To prove our analysis, we calculate the charge density difference (CDD) for the polarization-down with FM (bilayer MBT on $In_2Se_3$) and polarization-up with the FM (bilayer MBT on $In_2Te_3$) state (details can be found in Figure. S9). Indeed, we find there is electron transfer from bilayer MBT ($In_2Te_3$) to $In_2Se_3$ (bilayer MBT), which indicates hole (electron) doping for bilayer MBT. Here, the doping mechanism for the $In_2Se_3$ and $In_2Te_3$ substrate is different because of their different band alignment with bilayer MBT. As explained previously, the hole doping is more effective for promoting FM than the electron doping because of spin-orbit effects. This explains why $In_2Se_3$ is more effective for enhancing FM than $In_2Te_3$ since $In_2Se_3$ can perform hole doping for bilayer MBT. Another factor for enhancing FM is the carrier density. Our calculations indicate that the electron transfer from $In_2Se_3$ to bilayer MBT is 0.06e/f.u., while the value is only 0.01e/f.u. for $In_2Te_3$ case. This result also explains why $In_2Se_3$ is more effective for promoting FM for bilayer MBT compared with $In_2Te_3$ substrate.

## 6. Conclusions

We studied the relationship between carrier density and interlayer FM for bulk MBT and proposed a scheme to realize hole and electron transfer through control of the polarization states for the 2D ferroelectric substrate removing the need of chemical doping. We applied this idea to the bilayer MBT system and showed that the FM of bilayer MBT can be enhanced using $In_2Se_3$ and $In_2Te_3$ substrates with different polarization states. Moreover, possible topological phase transition between ZPQAH and QAH happens as the magnetic state changes. Our work provides a new strategy to realize QAH for even-layer MBT.

**Data availability statement**

The data that support the findings of this study are available upon reasonable request from the authors.

**Acknowledgement**

The research was supported by the US Department of Energy, Office of Science, Office of Basic Energy Sciences, Materials Sciences and Engineering Division (W.L., M.-H. D, F.R.), and by the U.S. Department of Energy (DOE), Office of Science, National Quantum Information Science Research Centers, Quantum Science Center (M.Y.). This research used resources of the Oak Ridge Leadership Computing Facility and the National Energy Research Scientific Computing Center, US Department of Energy Office of Science User Facilities.

**ORCID iDs**

Wei Luo https://orcid.org/0000-0002-2221-5913

Mao-Hua Du https://orcid.org/0000-0001-8796-167X

Fernando A. Reboredo https://orcid.org/0000-0003-1900-3888

Mina Yoon https://orcid.org/0000-0002-1317-3301

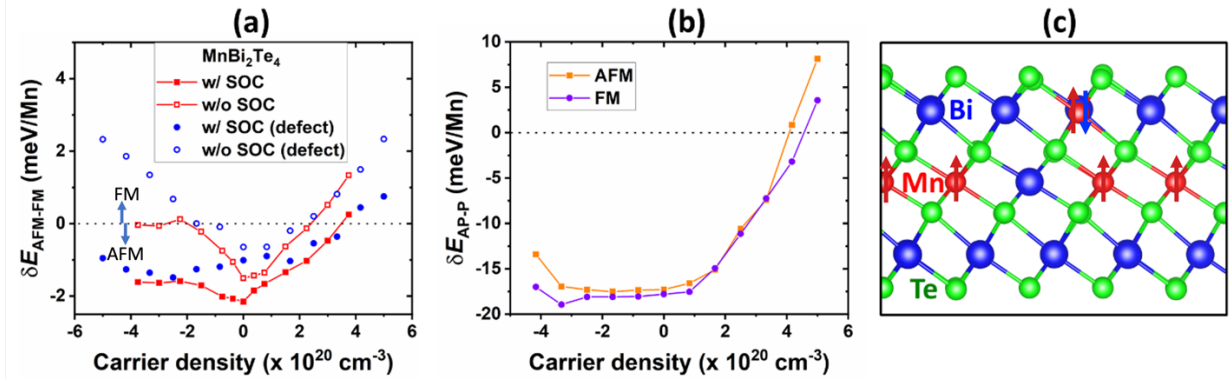

Figure. 1. (a) Calculated energy differences between AFM and FM as a function of the free carrier density in MBT. Free carriers are introduced by directly adding electrons or holes (red squares) or by creating shallow donors, $Bi_{Mn}^{+}$, or acceptors, $Mn_{Bi}^{-}$ (blue circles). Results obtained both with and without the SOC are shown. (b) Calculated energy differences between antiparallel and parallel alignments between the magnetic moments of a $Mn_{Bi}^{-}$ defect and its adjacent native Mn

atom (Mn$_{Mn}$) in both AFM and FM MBT as a function of the free carrier density without considering the SOC. (c) Crystal structure of a MBT SL with a pair of antisite-defects (Bi$_{Mn}^{+}$ and Mn$_{Bi}^{-}$). Red, blue, and green balls represent Mn, Bi, and Te ions, respectively. Red and blue arrows represent the direction of magnetic moment of Mn.

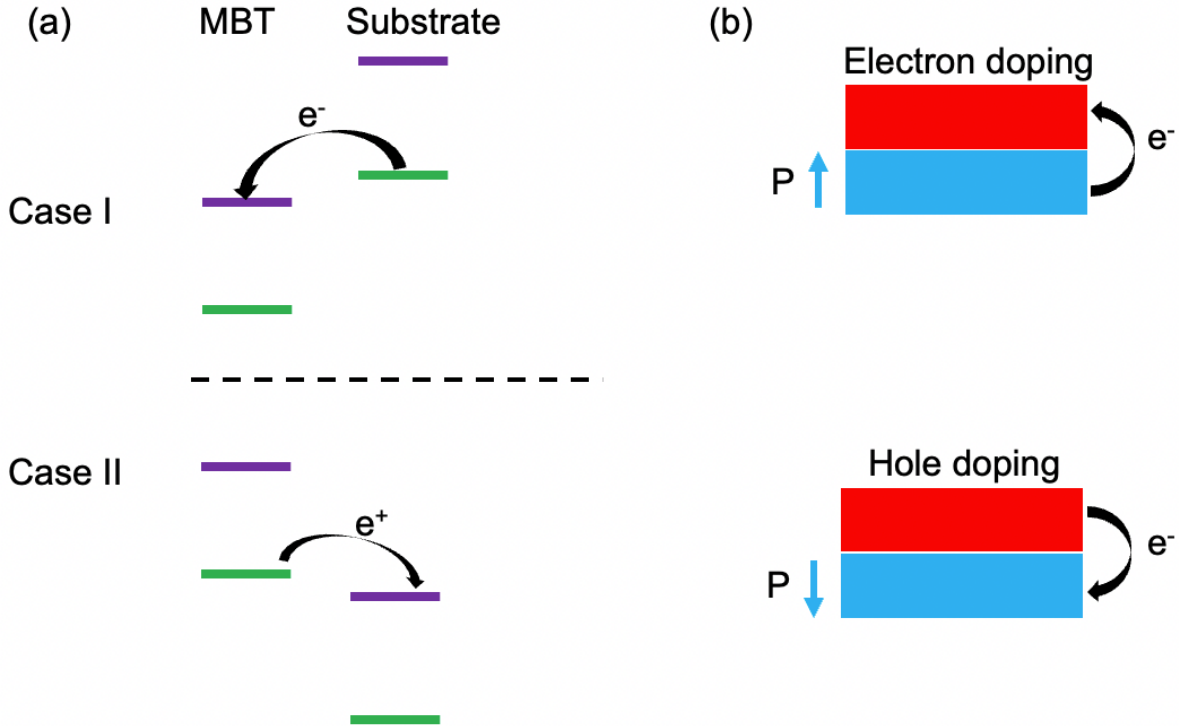

Figure. 2. (a) Type-III band alignment between the host material and the substrate, where purple and green lines represent the CBM and VBM, respectively. Case I (II) indicates that the substrate VBM (CBM) is higher (lower) than the host material CBM (VBM), resulting in electron (hole) doping of the host material. (b) The red and blue slabs represent the host material on the 2D ferroelectric substrate, where the direction of polarization on the substrate determines either hole of electron doping of the host material.

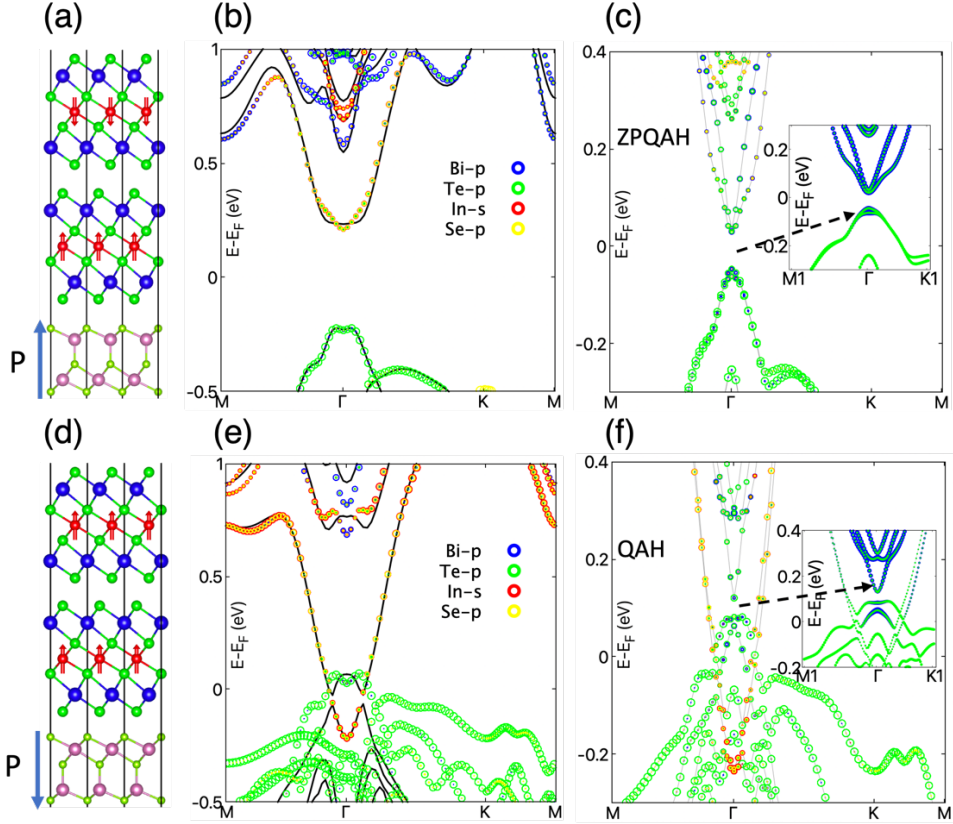

Figure. 3. (a) Crystal structure of heterostructure between bilayer MBT (AFM) and $In_2Se_3$ (polarization-up). The corresponding band structures are shown without (b) and with (c) SOC effects. (d) Crystal structure of heterostructure between bilayer MBT (FM) and $In_2Se_3$ (polarization-down). Note that the structural difference between polarization-up and polarization-down is the shift of Se atoms in the middle of the $In_2Se_3$ layer). The corresponding band structures are shown without (e) and with (f) SOC effects. Note that the black lines in (b) and (e) represent the spin down channel. M1 and K1 in the inset represent the points (0.25 0) and (0.1111, 0.1111) in BZ.

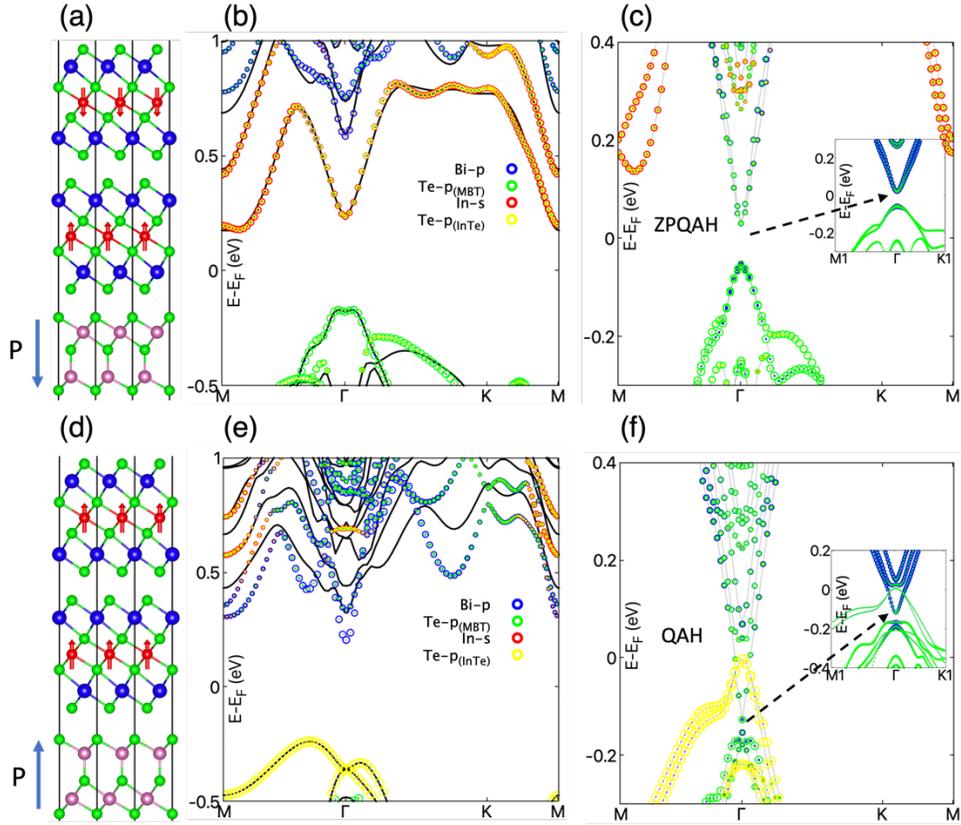

Figure. 4. (a) Crystal structure of heterostructure between bilayer MBT (AFM) and $In_2Te_3$ (polarization-down). The corresponding band structures are shown without (b) and with (c) SOC effects. (d) Crystal structure of heterostructure between bilayer MBT (FM) and $In_2Te_3$ (polarization-up). The corresponding band structures are shown without (e) and with (f) SOC effects.

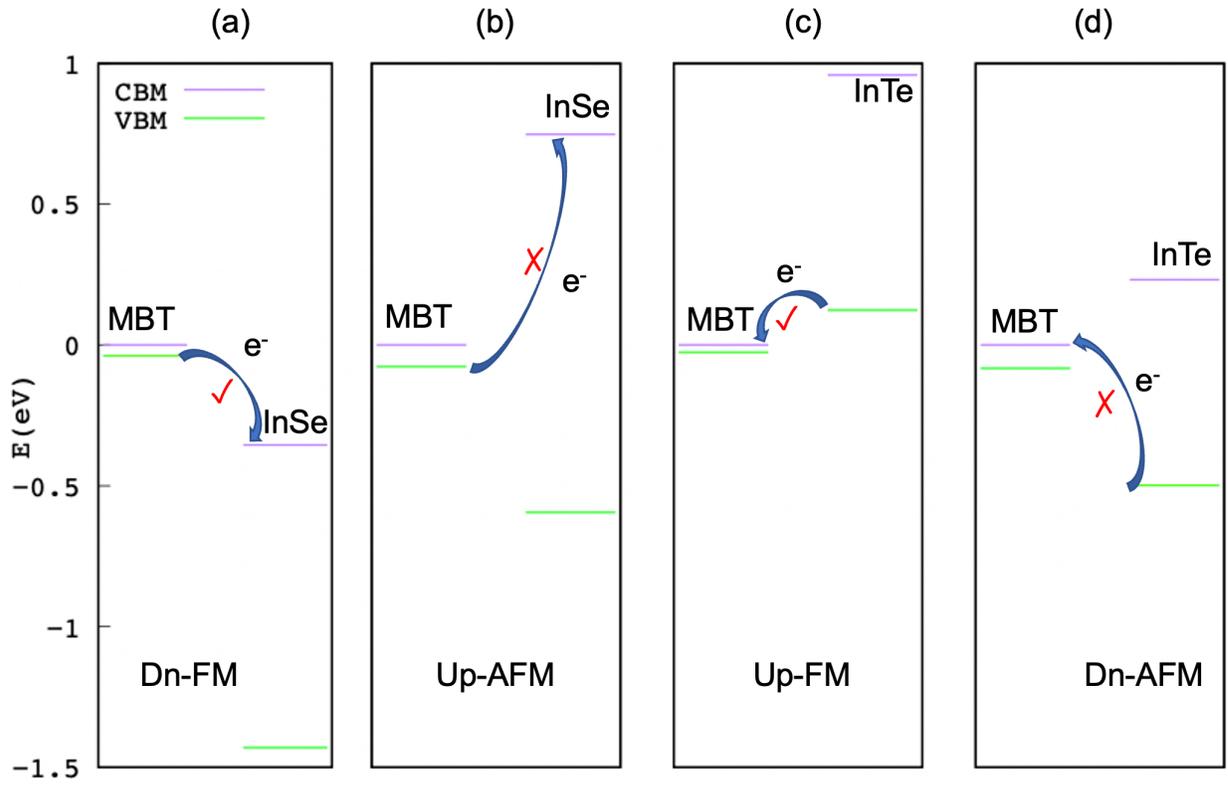

Figure. 5. Band alignment between MBT and $In_2Se_3$ ($In_2Te_3$) at Γ point. The CBM (purple line) of MBT is set to the reference point (zero energy) for each case.